\documentstyle[multicol,prb,aps,epsf]{revtex}
\begin{document}

\draft

\title{Coherent Transport through a Quantum Dot
       Embedded in an Aharonov-Bohm Ring}

\author{Kicheon Kang\cite{kang}}

\address{   Max-Planck-Institut f\"ur Physik Komplexer Systeme,
            N\"othnitzer Str. 38, D-01187 Dresden, Germany }

\date{\today}

\maketitle

\begin{abstract}
We study the coherent transport in multi-terminal mesoscopic
Aharonov-Bohm ring with a quantum dot embedded in an arm. 
Employing the Friedel sum rule for the effective 
single-particle levels in the
quantum dot, we explain some anomalous features which have been 
observed in the experiment.
We attribute these anomalies to the result of nontrivial 
quantum interference of the quantum dot with the attached ring.
Further, we propose a new feature of conductance oscillations
which can be a test for the validity of our model.
\end{abstract}
%
\begin{multicols}{2}
Resonant tunneling through a quantum dot is of considerable current 
interest (see e.g. ref.\onlinecite{kouwen97} and references therein).
The phase coherence of the resonant tunneling cannot be proved directly
in the ordinary conductance measurements through a quantum dot
because the conductance measures
only the magnitude of the transmission amplitude.
Yacoby {\em et al.}~\cite{yacoby95} reported the first experimental 
demonstration that transmission through a quantum dot has a coherent
component, using a two-terminal Aharonov-Bohm (AB) 
interferometer with a quantum dot embedded in one of its arm. In addition to
the observation of the coherence, they have found 
two other striking features which could not be understood well at first.
First, the phase of the AB oscillation changes abruptly by $\pi$ whenever
the conductance reaches its maximum. 
Second, the AB oscillations of successive
conductance peaks are in phase. It has been shown that the 
abrupt phase change at resonance can be understood in terms of the
phase rigidity enforced by the condition that the two terminal conductance
should be an even function of the external magnetic 
field\cite{yeyati95,hacken96,yacoby96}.
The second feature of {\em in phase} behavior has not been understood
well since neither integrable nor chaotic quantum dots are expected
to have generically the same phase between successive resonances.
Recently, Wu {\em et al}~\cite{wu98} suggested that the {\em in phase} behavior
originates from the fact that the resonant tunneling through the
whole system can be observed only when the phase shift introduced
by the resonant state of the dot coincides with the transmission phase 
of the reference arm.

In the two-terminal structure, a measurement of the transmission phase of
the quantum dot itself is not possible because of the phase rigidity
enforced by the micro-reversibility of the transmission 
coefficient~\cite{buttiker86}.
Recently, a modified four-terminal geometry has been adopted to measure
the transmission phase of the quantum dot\cite{schuster97}. 
Because the phase rigidity does not exist in this geometry,
they could observe continuous phase shifts of the AB oscillations
as a function of the plunger gate voltage on the quantum dot.
Within a simplified model that the coherent transmission can be described
by the sum of two direct paths, the phase evolution within a resonance
could be explained by the Breit-Wigner model for the quantum dot. 
On the other hand,
they observed two other striking phenomena which have not been 
understood by the Breit-Wigner model.
First, the AB oscillations of the successive resonances
are in phase again as in the two-terminal 
experiment. Second, there is sharp
phase drop, by $\pi$, at some
point between successive resonances, which is quite
different from the phase change at the peaks in the two-terminal experiment. 

In this paper, we address the problem of the coherent transmission
through a quantum dot embedded in a two-terminal and in a four-terminal
AB ring. By employing the 
Green's function method~\cite{fisher81,datta95} in
the tight binding model with the
Friedel sum rule for the quantum dot, we obtain some anomalous 
results which also have been observed in the experiments
such as {\em in phase} behavior and the inter-resonance phase drop.  
First, we confirm that the {\em in phase} behavior arises 
because the {\em out of phase}       
resonances in the quantum dot do not appear as a conductance
peaks due to the destructive interference, as Wu 
{\em et al.}~\cite{wu98} proposed.
Second, it is found
that the inter-resonance phase drop accompanies quite anomalous
AB oscillation, which cannot be described by a simple sum of two
direct paths. We show that multiple path contributions in the interference
are very important in the limit of small transmission probability and
closely related to the inter-resonance phase drop.
Further, we find an anomalous periodicity in the conductance oscillations,
with varying the value of the external magnetic flux.
From our result, we conclude that, in general, the transmission
through the quantum dot cannot be considered separately
from that of the whole
system containing the ring. The quantum
interferences lead to novel phenomena which cannot be understood in
terms of the quantum dot only.

The model we study is a multi-terminal AB ring where a quantum 
dot is embedded in one of its arm, as shown schematically in Fig.1.
We adopt a single channel model of spinless electrons. We use tight binding
representation with the hopping integral $t$ of which
magnitude is taken to be unity here. 
This model can be applied for
a ring where the ring is so narrow that
only a few 1D channels are included in tranmission. 
This is exactly the situation of
the experiments \onlinecite{yacoby95} and \onlinecite{schuster97}. 
Also we neglect
electron spin. The spin doesn't seem to play a major role
in the experiments
because it doesn't show an even-odd parity effect for the occupation
number of the quantum dot. 
The quantum dot is modelled by the barrier energy $E_B$ and multi-levels
in the site of the dot.
An equal spacing is assumed for the effective single particle energies
in the quantum dot. 
The periodicity comes from large charging energy,
which implies that the Coulomb interaction 
effects are being considered through
the effective single particle levels. 
That is, the energy levels in the quantum dot are
modelled as $E_l=E_0+l\Delta$ with $l=0,1,2\cdots$, 
where $\Delta\sim e^2/C$ with $C$ being
the capacitance of the quantum dot.
Due to the Friedel sum rule,
the increment in the occupation number $\delta n$ and the phase
shift $\delta\eta$ of the quantum dot levels are related by 
\begin{equation}
 \delta \eta = \pi\, \delta n .
\end{equation}
It should be noted that the sum rule is valid in spite of
the electron-electron interactions.
In a well confined quantum dot characterized
by a large value of $E_B$ in our model, 
the charge in the quantum dot is quantized by the charging energy 
so that we have $\delta n=1$ between adjacent levels.
Thus the adjacent levels have opposite parities with each other, 
which are denoted by solid lines and dotted lines drawn alternatively
in Fig.1(b).  This phase shift is taken
into account in the hopping matrix elements with the neighboring sites.
The magnetic flux $\Phi$ appears in the phase factor
$e^{\pm i\varphi}$ of the hopping integral, where 
$\varphi = 2\pi\Phi/N_s\Phi_0$  
with $\Phi_0=hc/e$ and $N_s$ being the elementary flux quantum and
the number of lattice sites, respectively.
The ring is connected to four reservoirs denoted by $L,R,a,b$ by the
coupling constant $t_{\alpha}$ ($\alpha=L,R,a,b$).
The coupling strength is characterized by the parameter
\begin{equation}
 \Gamma_{\alpha} = \pi |t_{\alpha}|^2 \rho_{\alpha}(\varepsilon_F) ,
\end{equation}
where $\rho_{\alpha}$ and $\varepsilon_F$ denote the density of states 
and the Fermi energy of the reservoir $\alpha$, respectively.

Owing to the relation between scattering amplitude
and the Green function~\cite{fisher81}, 
the transmission probability from the left lead to the right lead,
$T_{LR}$, can be related to the Green's function connecting the site
$1$ and the site $N$, $G_{1N}$:
\begin{equation}
 T_{LR} = 4\Gamma_L\Gamma_R |G_{1N}(\varepsilon_F)|^2 .
\end{equation}
Two-terminal system can be studied by taking $\Gamma_a=\Gamma_b=0$.
The two-terminal conductance is proportional to $T_{LR}$  
according to the Landauer formula.
In the four-terminal
geometry, $T_{LR}$ could be measured with an open circuit collector
($I_R=0$)~\cite{schuster97}. 
The Green function $G_{1N}$ is calculated by using the standard
Green function technique in the presence of multi-terminal
leads~\cite{fisher81,datta95}.

Fig.2(a) displays the transmission probability and its phase ($\theta$)
of the AB oscillation as a function of the lowest
dot level $E_0$. In the experiment, $E_0$ can be controlled by the 
external plunger gate on the quantum dot. As observed in the experiments,
periodic conductance oscillation due to the
charging energy is shown in this figure. 
The asymmetry of the peaks is the result of the interference with the 
upper arm.
Surprisingly, the period of the oscillation is not $\Delta$ but $2\Delta$.
This implies that the oscillation period corresponds to
adding charge $2e$ to the quantum dot, not $e$ as in the ordinary
Coulomb blockade oscillations. 
In the two-terminal structure, the conductance should be an
even function of the external flux, which allows only abrupt phase
change of $\pi$~\cite{buttiker86}. 
It should be noted that
there are two types of abrupt phase changes. One occurs at resonance and the
other does at some point between adjacent resonance peaks.
With two types of phase changes in a period, 
the every peak has the same phase in AB
oscillations as observed in the experiments\cite{yacoby95,schuster97}. 
As a result of the interference with the reference arm,
only {\em in phase} resonances through the quantum dot appear
as conductance peaks, while the {\em out of phase} resonances don't
give rise to conductance peaks because of destructive quantum 
interference. This explains why the conductance peaks in
the experiment should have the same phase in AB oscillations.

AB oscillations are more closely inspected in Fig.2(b).
In Fig.2(b), the transmission probability as a function of the flux
is displayed for several values of $E_0$. One can see that
the parity of AB oscillations is changed twice in a period
as mentioned above. The phase change at the conductance
peaks is now well understood
from the previous studies~\cite{yeyati95,hacken96,yacoby96}.
Further, we find that the phase change between the peaks accompanies
quite anomalous AB oscillation. That is, at the point of
inter-resonance phase change ($X$), the transmission amplitude
is zero for almost every value of the external flux. It means that
coherent transmission is nearly absent at this point.
This result is quite similar to the experimental observation of
\cite{schuster97} that the inter-resonance phase drop accompanies
zero amplitude of the AB oscillation.  

In Fig.3, transmission probability and its phase of AB oscillation
in the four-terminal geometry is displayed
as a function of $E_0$. While a continuous phase shift has been 
observed in the experiment~\cite{schuster97}, the phase rigidity still exists
in our treatment
because net current flow through the other reservoirs $a$, $b$ is not allowed
in this formulation. It has been shown that the phase
rigidity is preserved even in the presence of inelastic processes
through the other leads if the net current flow is zero through these 
leads~\cite{yeyati95}.
In our treatment, the behavior
of AB phase is same as that of two-terminal ring,
with reduced transmission
probability. 
The reduction of the probability comes from the fact that the additional
reservoir plays a role of inelastic scattering center.~\cite{buttiker88}
The AB oscillation patterns which are not displayed here
are same with those of two-terminal interferometer (Fig.2(b)) 
with reduced magnitude of the oscillations.
Looking into the AB oscillations displayed in Fig.2(b) again, 
one can find that the magneto-conductance curves are far from sinusoidal 
of the period $\Phi_0$
at low transmission region. 
This means that multiple path interference is important
especially at low transmission region (see {\em C, D, X} in Fig.2(b)).
In contrast, near the resonance,
the multiple path contributions are relatively small and the $\Phi_0$
period due to direct two paths is dominant (see {\em A, R, B} in the
Fig.2(b)). This suggests that multiple path contributions should not
be neglected in the small transmission limit, and it requires more
careful analysis.
In analyzing their experimental results, the authors of
\onlinecite{schuster97} used a simple model of sum of two direct paths,
on the basis of their observation that there is no
higher order harmonics in the 
AB oscillation with period $\Phi_0/n$ ($n>1$). While it seems valid in 
describing the phase evolution around the peaks, it is still 
questionable whether this argument is correct in the limit of small amplitude
of AB oscillation. Though our treatment is not complete because of
the restriction in the allowed phase values, numerical results
indicate at least that multiple path interference
cannot be neglected at low transmission limit. Further, multiple
path contributions are closely related to the inter-resonance phase drop.
At the point of inter-resonance phase drop ({\em X} of Fig.2(b)), 
effects of multiple 
path interferences are rather drastic, which lead to quite anomalous
AB oscillation pattern. 

In Fig.4, we display $T_{LR}$ as a function of $E_0$ for several
values of the external flux quantum. Interestingly, the periodicity
varies as $2\Delta$ $\rightarrow$ $\Delta$ $\rightarrow$ $2\Delta$
$\rightarrow$ $\Delta$ with increasing the value of the flux.
For $\Phi=\Phi_0/2$, the locations of the peaks are shifted by 
$\Delta$ compared to the zero flux case. This is because the ring acquires 
AB phase $\pi$ due to the flux, 
so the {\em in phase} resonances and {\em out of phase}
resonances in the quantum dot are reversed.
When the periodicity is $\Delta$
($\Phi=\Phi_0/4$, $3\Phi_0/4$), the conductance peaks are no longer in phase,
and the positions of the peaks do not coincide with the resonance of 
the quantum dot. This is also the result of interference.
We suggest that the validity of the model presented in this paper
can be tested experimentally by investigating the feature in
Fig.4.

In conclusion, we have investigated the coherent transmission
in two-terminal and in four-terminal mesoscopic
Aharonov-Bohm ring with a quantum dot embedded in an arm. 
Employing the Friedel sum rule for the 
effective single-particle levels in the
quantum dot, we have explained some anomalous features which have been 
observed in the experiment. 
We have discussed these anomalous features in relation to the nontrivial 
quantum interference of the quantum dot with the attached ring.
Further, we have proposed a new feature of conductance oscillations
which can be a test for the validity of our model.

The author thanks H. Schanz for his critical reading of this manuscript,
and P. Fulde for his hospitality during his stay in the MPI-PKS.
This work has been supported by the 
KOSEF and in part by the Visitors Program of the MPI-PKS.


%
%
\begin{figure}[t]
 \vspace*{-2cm}
 \epsfxsize=3.5in
 \epsffile{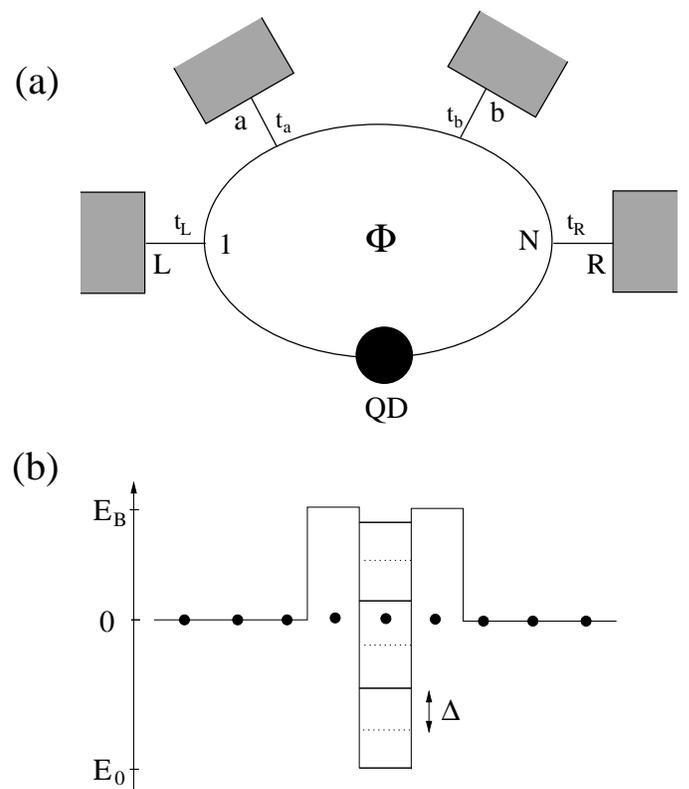}
 \vspace*{-2cm}
  \caption{ (a) Schematic diagram of the Aharonov-Bohm ring with 
                a quantum dot coupled to four reservoirs denoted by
                $L$, $R$, $a$ and $b$ with coupling constant $t_{\alpha}$
                ($\alpha=L,R,a,b$).
            (b) The on-site energies in the tight binding model. The
                quantum dot is modelled by the barrier energy $E_B$
                and periodic multi-level energies with
                its lowest level $E_0$ and spacing $\Delta$, 
                respectively.
                The values of their phase shifts are given by
                $0$ (solid line) and $\pi$ (dotted line) alternatively
                due to the Friedel sum rule.
	   }
\end{figure}
\begin{figure}[t]
 \vspace*{-2cm}
 \epsfxsize=3.5in
 \epsffile{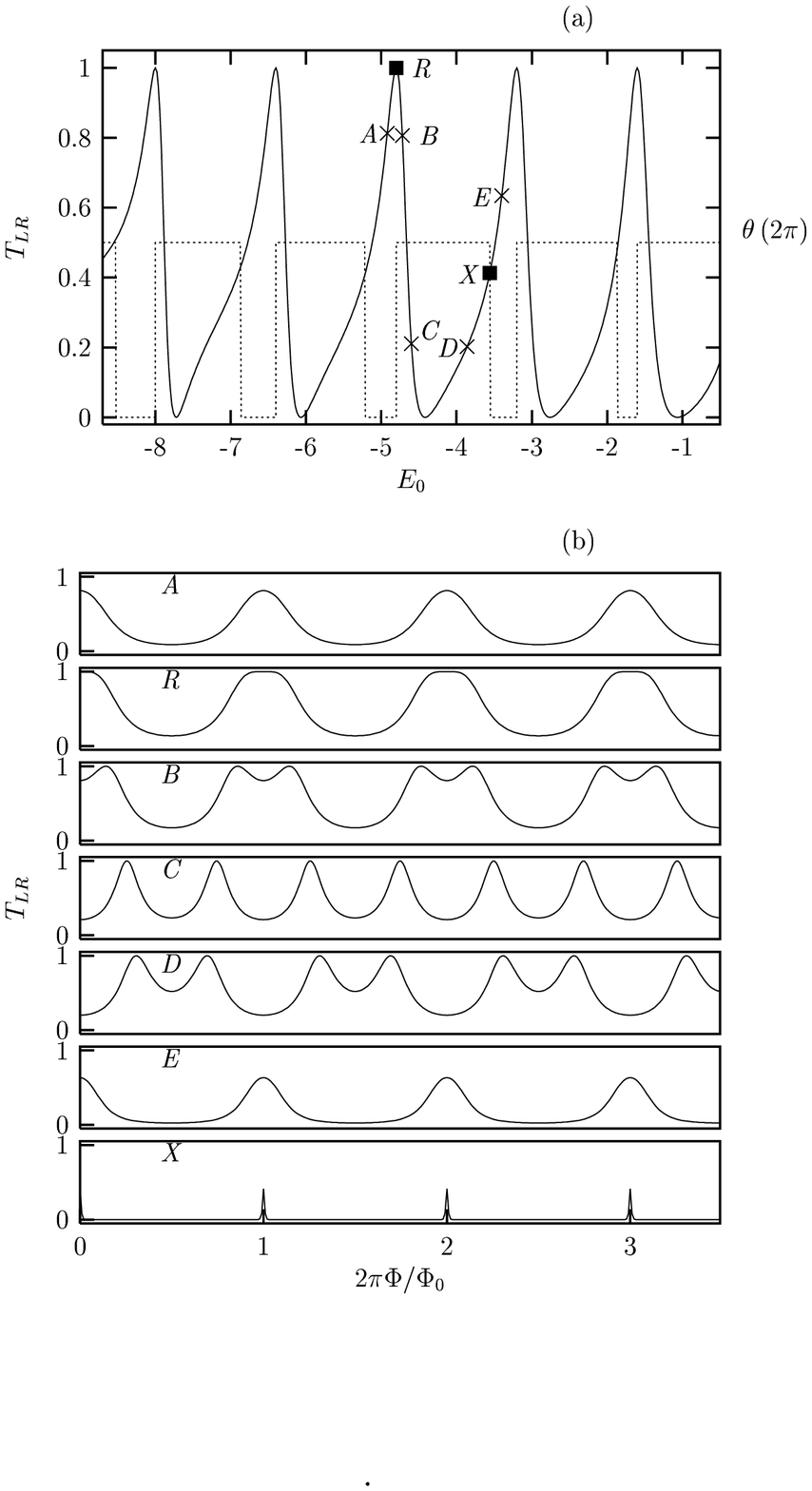}
 \vspace*{-2cm}
  \caption{ 
    Transmission probability and phase of AB oscillation
    as a function of the
    lowest energy of the dot level $E_0$ 
    in the two-terminal ring. The parameters
    used for the calculations are $\Delta=0.8$, $\Gamma_L=\Gamma_R=0.1$,
    $\Gamma_a=\Gamma_b=0$ and $E_B=4.0$ in unit of $t$. 
    $E_0$ is also normalized in unit of $t$.
    12 levels in the quantum
    dot are taken into account in the calculation. 
    (a) Transmission probability (solid line) in the 
    absence of the external magnetic flux and its phase
    of the AB oscillation (dashed line).
    (b) AB oscillations of the transmission probability
    for several values of $E_0$ marked
    as {\em A, R, B, C, D, E} and {\em X} in (a).
   }
\end{figure}
\begin{figure}[t]
 \vspace*{-2cm}
 \epsfxsize=3.5in
 \epsffile{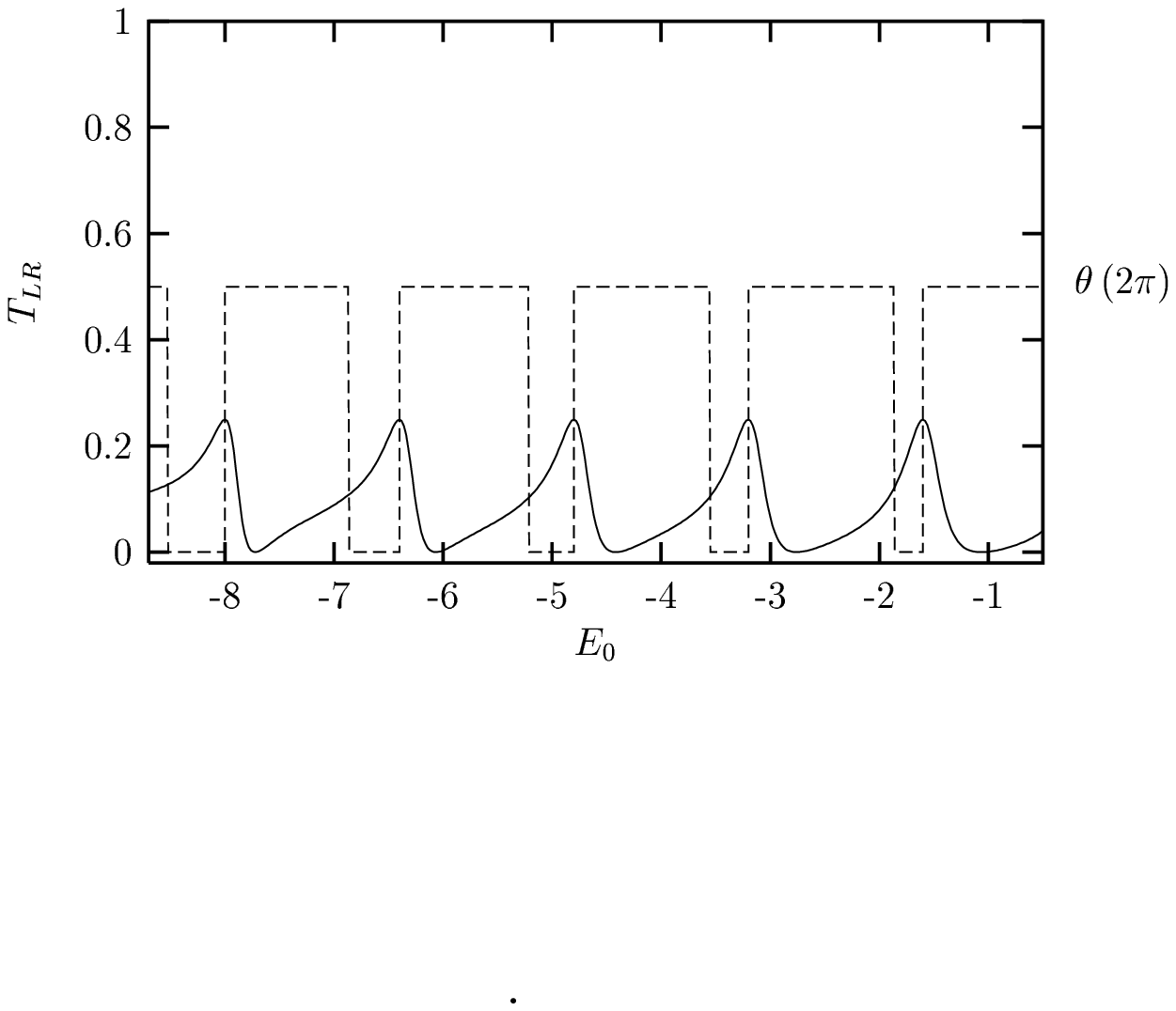}
 \vspace*{-7.5cm}
  \caption{ 
    Transmission probability (solid line) from the left ($L$) to the
    right ($R$) lead with its phase (dashed line)
    in the four-terminal geometry. The
    coupling strength is given by $\Gamma_L=\Gamma_R=\Gamma_a=\Gamma_b=0.05$,
    in unit of $t$.
    Other parameters are same with those in Fig.2.}
\end{figure}
\begin{figure}[t]
 \vspace*{-2cm}
 \epsfxsize=3.5in
 \epsffile{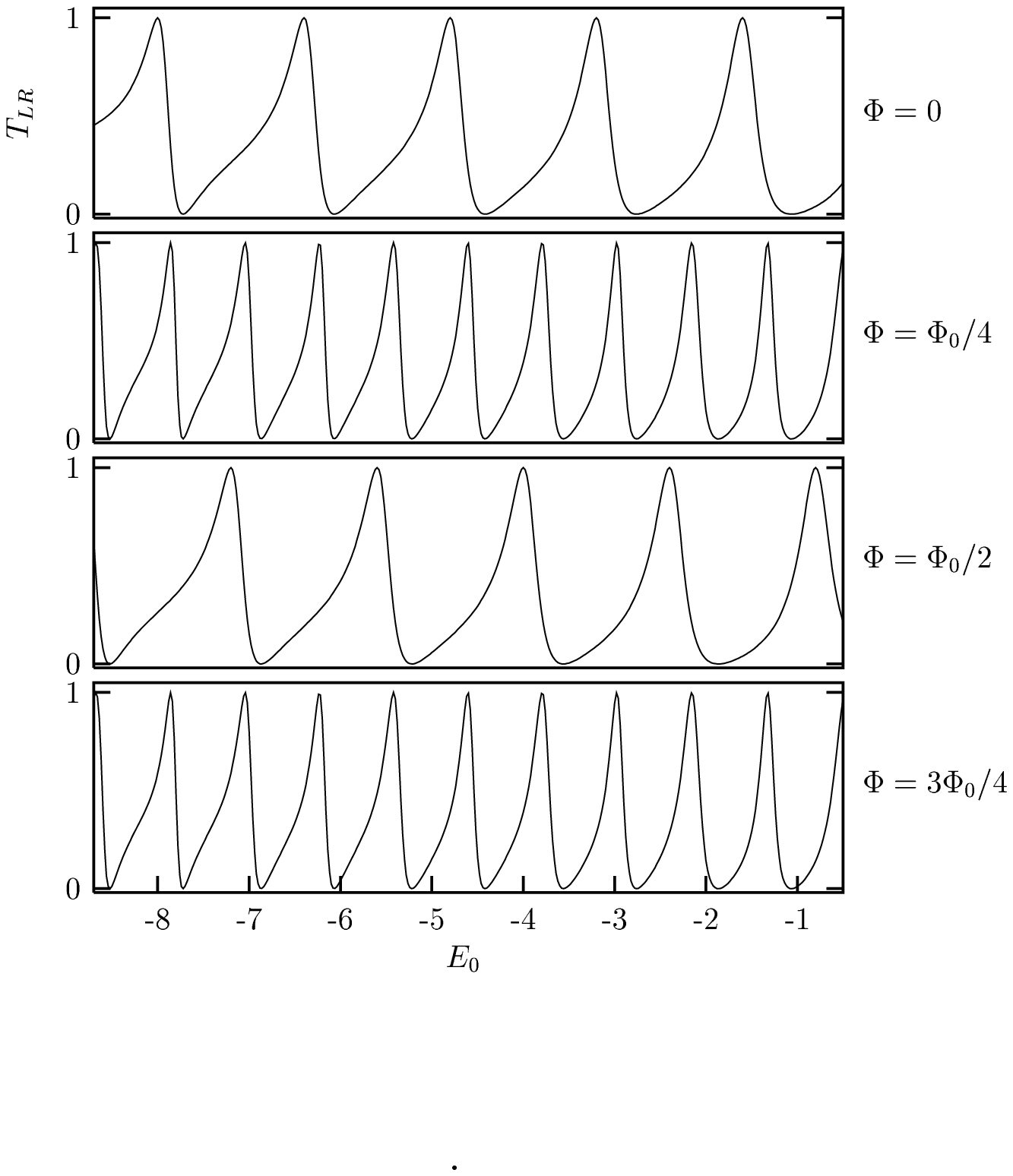}
 \vspace*{-5cm}
  \caption{
    Transmission probability as a function of $E_0$ in the two-terminal
    geometry for several values of the external flux. Other parameters
    are same with those in Fig.2.} 
\end{figure}
\end{multicols}
\end{document}